\documentclass{emulateapj}
\usepackage{ulem}
\usepackage{longtable}
\usepackage{graphicx}
\input{epsf.sty}
\input{psfig.sty}

\shorttitle{Jet precessing in Sw J1644+57}

\begin{document}
\title{Frame-dragging, disk warping, jet precessing, \\
and dipped X-ray lightcurve of Sw J1644+57}
\author{Wei-Hua Lei$^{1,2}$, Bing Zhang$^{2,3}$, He Gao$^2$}
\altaffiltext{1}{School of Physics, Huazhong University of Science and Technology, Wuhan, 430074, China}

\altaffiltext{2}{Department of Physics and Astronomy, University of Nevada Las Vegas, 4505 Maryland Parkway, Box 454002, Las Vegas, NV 89154-4002, USA. Email: leiwh@physics.unlv.edu; zhang@physics.unlv.edu}

\altaffiltext{3}{The Kavli Institute for Astronomy and Astrophysics and Department of Astronomy, Peking University, Beijing 100871, China}

\begin{abstract}
The X-ray transient source Sw J1644+57 recently discovered by Swift is believed to be triggered by tidal disruption of a star by a rapidly spinning supermassive black hole (SMBH). For such events, the outer disk is very likely misaligned with respect to the equatorial plane of the spinning SMBH, since the incoming star before disruption most likely has an inclined orbital plane. The tilted disk is subject to the Lense-Thirring torque, which tends to twist and warp due to the Bardeen-Petterson effect. The inner disk tends to align with the SMBH spin, while the outer region tends to remain in the stellar orbital plane, with a transition zone around the Bardeen-Petterson radius. The relativistic jet launched from the spinning SMBH would undergo precession. The 5-30 day X-ray lightcurve of Sw J1644+57 shows a quasi-periodic (2.7-day) variation with noticeable narrow dips. We numerically solve a warped disk and propose a jet-precessing model by invoking a Blandford-Znajek jet collimated by a wind launched near the Bardeen-Petterson radius. Through simulations, we show that the narrow dips in the X-ray lightcurve can be reproduced for a range of geometric configurations. From data we infer that the inclination angle of the initial stellar orbit is in the range of $10^{\circ}-20^{\circ}$ from the SMBH equatorial plane, that the jet should have a moderately high Lorentz factor, and that the inclination angle, jet opening angle, and observer's viewing angle are such that the duty cycle of the line-of-sight sweeping the jet cone is somewhat less than 0.5.
\end{abstract}

\keywords{ accretion, accretion disks--black hole physics--magnetic fields}

\section{Introduction}
The discovery of the hard X-ray transient event Swift J16449.3+573451 (``Sw J1644+57'' hereafter, Burrows et al. 2011) by the Swift satellite (Gehrels et al. 2004) has stimulated a great interest in studying jets launched from tidal disruption events (TDEs). The long variability time scale $\delta t \sim 100$ s (Burrows et al. 2011) and its location near the center of a $z=0.354$ host galaxy (Levan et al. 2011) link Sw J1644+57 to a super-massive black hole (SMBH) (Burrows et al. 2011; Bloom et al. 2011). The sharp onset and gradual fade-away of X-ray flux refer to tidal disruption of a star by a dormant SMBH (Bloom et al. 2011; Burrows et al. 2011). The super-Eddington X-ray luminosity (Burrows et al. 2011), bright radio afterglow (Zauderer et al. 2011), as well as a stringent historical X-ray flux upper limit suggest that a relativistic jet is launched from a SMBH during the TDE, which is not expected in most previous TDE studies (e.g. Kobayashi et al. 2004; Strubble \& Quataert 2010; Lodato \& Rossi 2011, but see Lu et al. 2008; Gao et al. 2010; Giannios \& Metzger 2011). Modelling the emission of Sw J1644+57 suggests that the jet is highly ``particle starved'' (Burrows et al. 2011), favoring a magnetically launched jet, likely launched via the Blandford-Znajek (1977, hereafter BZ) mechanism (Lei \& Zhang 2011), which extracts the spin energy of the BH through a magnetic field connecting the BH event horizon and a remote astrophysical load. From observational constraints, Lei \& Zhang (2011) found that the SMBH in Sw J1644+57 carries a moderate to high spin. 

For such events, since the star initially has no knowledge about the BH spin orientation before being disrupted and accreted, it is most likely that the initial stellar orbit is mis-aligned with the equatorial plane of the spinning BH. A natural expectation is that at least the outer part of the accretion disk is also misaligned. The tilted disk surrounding a spinning BH is subject to the Lense-Thirring (hereafter LT, Lense \& Thirring 1918) torque. The combined action of the LT torque and the internal viscosity of the accretion disk would lead to a twisted and warped disk, with the inner part of the disk bent towards the BH equatorial plane due to the frame-dragging effect. This is known as the Bardeen-Petterson (hereafter BP) effect (Bardeen \& Petterson 1975). For a fully developed BP disk, the inner part of the disk tends to be aligned with the BH equatorial plane, while the outer part of the disk remains aligned with the original stellar orbital plane. The transition radius between these two regimes is the BP radius. The warped disks have been directly observed by water maser observations in NGC 4258 (Miyoshi et al. 1995; Neufeld \& Maloney 1995; Herrnstein, Greenhill \& Moran 1996) and the Circinius galaxy (Greenhill et al. 2003). The apparent lack of correlation between the direction of the radio jets emanating from active galactic nuclei (AGN) and the plane of the host galaxy disks (Kinney et al. 2000; Schmitt et al. 2002) can also be explained by disk warping. 

Besides this theoretical motivation, some curious observational facts also point towards the possibility of a warped disk. Burrows et al. (2011) reported a rough period of 230 ks ($\sim$ 2.7 days) in the long-term X-ray lightcurve, although with a $< 3 \sigma$ significance. Saxton et al. (2012) gave a more thorough analysis and derived a similar period with higher significance. More importantly, a closer inspection suggests that the late X-ray lightcurve show dips. A natural mechanism to account for these dips would be a precessing jet, possibly caused by a warped BP disk (see also Saxton et al. 2012). Stone \& Loeb (2012) discussed the similar topic, but didn't pay attention to these peculiar observational effects. They led to the conclusion that the stellar orbital plane is almost aligned with the BH equatorial plane due to the lack of observational evidence of jet precessing.

In this paper, we develop a model to interpret the quasi-periodic feature and the narrow dips observed in the lightcurve of Sw J1644+57 within the framework of the BP mechanism. In Section 2, we first apply a mathematical method, i.e. Stepwise Filter Correlation (SFC) method (Gao et al. 2012), to study the X-ray lightcurve data of Sw 1644+57. We confirm that a $\sim 2.7$-day quasi-periodic signal exists in the late lightcurve after the initial flaring phase.  We then develop a numerical model in Section 3 to study disk warping and jet precessing. In Section 4, we apply the observational data to constrain the model parameters, and found several parameter sets that can roughly reproduce the observations. Our conclusions are presented in Section 5.

\section{Observational evidence of jet precession}

\begin{figure}[htc]
\center
\includegraphics[width=7cm,angle=0]{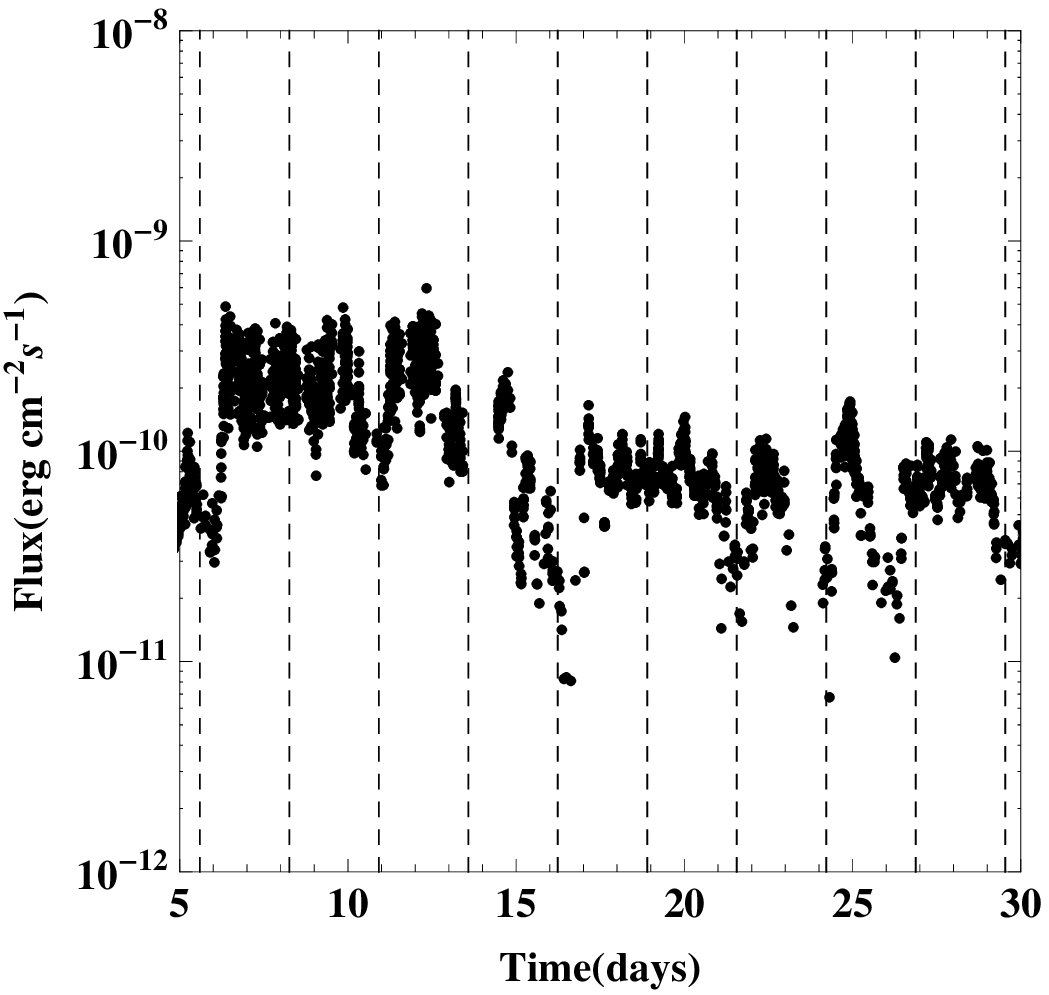}\\
\includegraphics[width=7cm,angle=0]{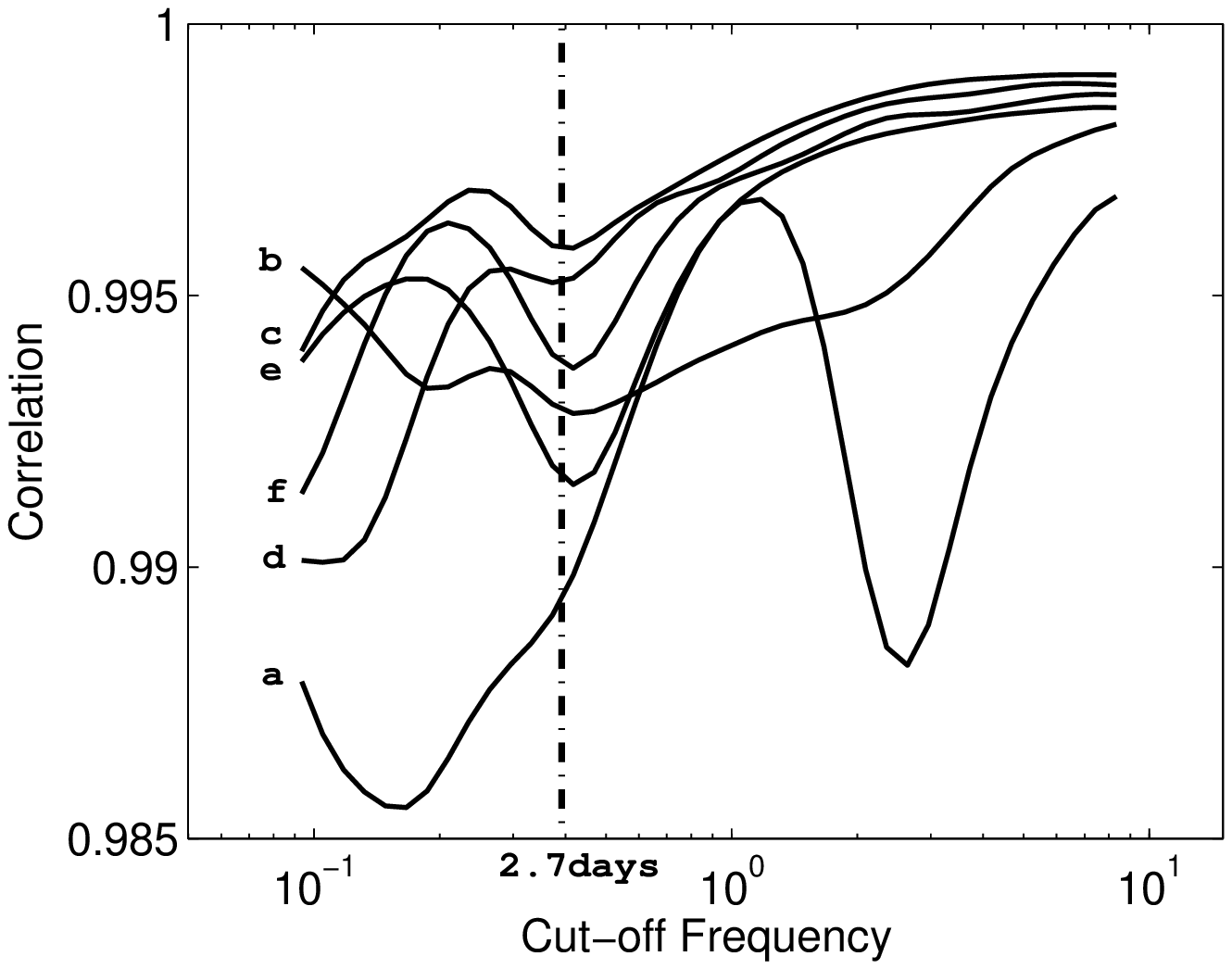}
\caption{Upper panel: Swift XRT light curve of Sw J1644+57 from 5-30 days, which show clear dips. The dashed line shows the 2.7-day period. Data is from Burrows et al. (2011). Bottom panel: the stepwise filter correlation (SFC) analysis of the X-ray lightucrve following Gao et al. (2012). The curve (a) is the result for the entire lightcurve, curve (b), (c), (d), (e) and (f) are the analyses by manually removing the first 2, 5, 10, 15, and 20 days of the data, respectively. A dip corresponding to $\sim$ 2.7 days is evident once the first 2-day of data is removed.}
\label{fig1}
\end{figure}

Two pieces of evidence hint towards the possibility of a precessing jet in Sw J1644+57. The first piece is the noticeable dips in the XRT lightcurve at late times. The upper panel of Fig.\ref{fig1} shows the XRT lightcurve of Sw J1644+57 after 5 days since the first trigger. Some deep dips are clearly visible. The dashed vertical lines mark the 2.7-day period (see more below). One can see that the dips roughly track this period. Dips are not easy to explain within models invoking emission, but would be naturally interpreted within the frame work of jet precession where the line of sight is allowed to move away from the emission beam. 

The second piece of evidence is the rough 2.7-day periodicity, which was pointed out by Burrows et al. (2011) and confirmed by Saxton et al. (2012). We investigate this independently using a new method, Stepwise Filter Correlation (SFC) method, recently developed in our group (Gao et al. 2012). The SFC method is a signal processing algorithm which stepwisely filters signals above a frequency, and looks for correlation between the lightcurves of adjacent filtered lightcurves. If there is a spectral component around a particular frequency, the lightcurves before and after filtering this particular frequency can be very different, leading to a dip in the correlation curve. The details of this method and its application to GRB lightcurves are presented in Gao et al. (2012). Here we apply the method to the XRT lightcurve of Sw J1644+57. We first perform an analysis to the entire lightcurve (curve (a) of lower panel of Fig.\ref{fig1}). Since the early lightcurve displays many erratic, bright X-ray flares, the signal is dominated by these high-frequency components, and the 2.7-day component cannot be identified. These flares may be related to fallback of the stellar debris with high and fluctuant accretion rate (Cannizzo et al. 2011). If we remove the early data (say, the first two days), the 2.7-day dip clearly shows up (curve (b) of lower panel of Fig.\ref{fig1}). We then gradually remove more and more early data points and everytime redo the SFC analysis (curves (c-e) of the lower panel of Fig.\ref{fig1}). The 2.7-day dip is always there. Our analysis suggests that, consistent with the results of Burrows et al. (2011) and Saxton et al. (2012), that the 5-30 day X-ray lightcurve of Sw J1644+57 has a quasi-periodic signal with a period roughly 2.7 days.


\section{Warped accretion disk and Jet precession}
\begin{figure}[htbp]
\centerline{\includegraphics[width=70mm,height=70mm]{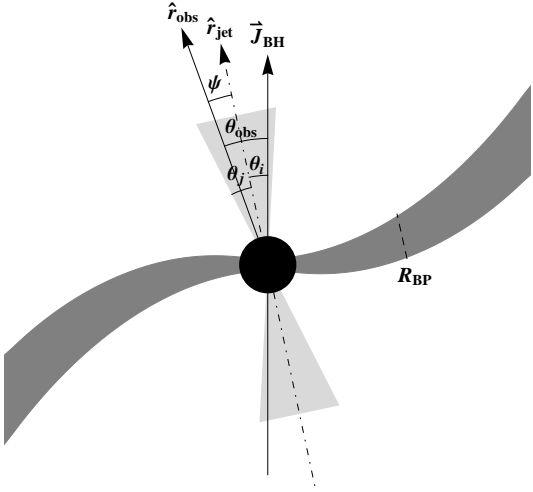}}
\caption[]{The sketch of the model. A star is tidally disrupted as it gets close to a dormant SMBH. The star orbit is generally inclined with the SMBH spin vector $\mathbf{J}_{\rm{BH}}$. After disruption, an accretion disk form, and is warped due to the Bardeen-Petterson effect. The jet is expected to precess with an angle $\theta_{\rm{i}}$ around the BH spin axis. For a jet with half opening angle $\theta_{\rm{j}}$ and an observer located at $\theta_{\rm{obs}}$ from the BH spin axis, the angle $\psi$ between the observer's line of sight $\hat{r}_{\rm{obs}}$ and jet axis $\hat{r}_{\rm{jet}}$ is changing with time. The line of sight can exit the jet cone at certain phases, leading to dips in the lightcurve. The Bardeen-Petterson radius $R_{_{\rm BP}}$ is marked. The jet direction is defined by the disk normal direction around $R_{_{\rm BP}}$.} \label{fig2}
\end{figure}

In simple physical terms, disruption of a star occurs when it comes to a SMBH closer than the tidal disruption radius $R_{\rm{T}}$, which is determined by demanding that the mean density of space volume enclosed by $R_{\rm{T}}$, i.e., $M_{\bullet}/(4\pi R_{\rm{T}}^3/3)$, is equal to the density of the star, where $M_{\bullet}$ is the mass of the BH. The tidal disruption radius is then given by
\begin{eqnarray}
R_{\rm{T}} \simeq (\frac{M_{\bullet}}{M_*})^{1/3} R_* & & \simeq  7 \times 10^{12} m_*^{-1/3} r_* M_{\bullet,6}^{1/3} cm \nonumber \\
&& \simeq 47 R_{\rm{g}} m_*^{-1/3} r_* M_{\bullet,6}^{-2/3},
\end{eqnarray}
where $m_*=M_*/M_{\sun}$, $r_* = R_*/R_{\sun}$, $M_{\bullet,6}=M_{\bullet}/10^6M_{\sun}$ and $R_{\rm{g}}=G M_{\bullet}/c^2$. For a solar-like star ($M_*=M_{\sun}$ and $R_*=R_{\sun}$) disrupted by a $2\times 10^6M_{\sun}$ BH, the tidal disruption radius is approximately $30 R_{\rm{g}}$ from the BH.

Stellar disruption occurs if $R_{\rm{ms}} \leq R_{\rm{P}} \leq R_{\rm{T}}$, where $R_{\rm{P}}$ is the periastron distance from the star, and $R_{\rm{ms}}$ is the marginally stable orbit (Bardeen et al. 1972). The star should be swallowed whole instead of being disrupted if it goes inside $R_{\rm{ms}}$. We define $\kappa \equiv R_{\rm{P}}/R_{\rm{T}}$. 

We idealize the disk as extending from $R_{\rm{ms}}$ to the circularization radius $R_{\rm{out}} \simeq \eta R_{\rm{P}}$, where $\eta \simeq 2$. In our calculation, the typical values of the parameters are $\eta=2$, $m_*=1$, $r_*=1$,and $\kappa=1$.

The dragging of the inertial frame (frame dragging) produced by a Kerr black hole causes precession of a particle if its orbital plane is inclined with respect to the equatorial plane of the black hole. This effect is known as Lense-Thirring (LT) precession. The precession angular velocity $\Omega_{\rm{LT}}$ is given by (e.g. Wilkins 1972)

\begin{equation}
\Omega_{\rm{LT}}(R)= \frac{2G}{c^2} \frac{J_{\bullet}}{R^3},
\end{equation}
where $J_{\bullet}=a_{\bullet} G M_{\bullet}^2/c$ is the BH angular momentum, and $a_{\bullet}$ is the BH spin parameter.

The precession period $\tau_{\rm{p}}$ can be estimated as
\begin{equation}
\tau_{\rm{p}} = 2\pi /\Omega_{\rm{LT}}
\label{eq_taup}
\end{equation}
For a BH with mass $2\times 10^6 M_{\sun}$ and spin $a_{\bullet}=0.9$, we have $\tau_{\rm{p}} \simeq 2.7$ days at $R=19R_{\rm{g}}$. 

When an accretion disk does not align with the equatorial plane of the BH, i.e., when the angular momentum of the accretion disk $\mathbf{L}$ is misaligned with respect to the direction of $\mathbf{J_{\bullet}}$, the LT effect causes a precession effect that twists the disk plane due to the coupling of $\mathbf{J_{\bullet}}$ with $\mathbf{L}$. The torque tends to align the angular momentum of the matter in the disk with that of the BH, thus causing the inclination angle between the angular momentum vectors to decrease with decreasing distance from the BH. This is the BP effect (Bardeen \& Petterson 1975), which is the combination of the LT effect and the internal viscosity of the accretion disk. The effect is more prominent in the inner part of the disk due to the short range of the LT effect. The outer part of the disk tends to remain in the orientation defined by the original stellar orbit.

Considering that the disk may be thick, Stone \& Loeb (2012) suggested that the disk is not subject to the BP warping effect. This conclusion is based on an inferred large accretion rate, and a weak magnetic field viscosity as revealed in a GRMHD simulation (Fragile et al. 2007). Motivated by the observational evidence discussed in Section 2, in this paper we suggest that the disk can be thin, and the BP effect indeed plays a role, at least after the initial flares when the accretion rate drops significantly. To justify this assumption, we argue that the dynamical viscosity and magnetic stress in the disk can be larger than what is invoked by Fragile et al. (2007). For relativistic twisted disk around a Kerr BH, Zhuravlev \& Ivanov (2008) found that the BP effect can take place if the viscosity parameter $\alpha$ is sufficiently large. More importantly, the absolute accretion rate of Sw J1644+57 is not well constrained from the data. The previous estimates (e.g. Burrows et al. 2011; Lei \& Zhang 2011) were derived by normalizing the total X-ray fluence with the total mass of the star, and use the flux to derive the accretion rate under the assumption that all the mass of the star is accreted into the black hole. For an assumed stellar mass $\sim 0.1 M_\odot$, the peak accretion rate is around $10^{-6} M_{\sun} s^{-1}$ (Lei \& Zhang 2011; Shao et al. 2011), and the late time (when dips show up) accretion rate is $\sim 10^{-8} M_{\sun} s^{-1}$, about 2 orders lower. This corresponds to the Eddington luminosity of a $10^7M_{\sun}$ BH with disk efficiency of about 10\%. However, the real accretion rate can be a factor of a few to 10 lower than this value. For example, the accreted star may be even less massive. Also according to recent MHD simulations by Yuan et al. (2012a, 2012b), a hot accretion flow would have a strong mass outflow. The net accretion rate in the inner disk would be then significantly reduced. Considering these factors, it is possible that the condition $H/R<\alpha$ is satisfied, where $H$ is half thickness of the disk, and $\alpha$ is the viscosity parameter. We could therefore assume that at least the inner disk is thin after the flare phase, so that the BP effect would come into play to warp the disk.

For a warped disk, we follow the formalism of Pringle (1992). There are two viscosity parameters $\nu_1$ and $\nu_2$, where $\nu_1$ is the standard shear viscosity in a flat disk, and $\nu_2$ is the viscosity associated with the vertical shear motion describing the diffusion of warping distortion through the disk. In order to study the BP effect, We re-write Pringle's equation by adding an effective coupling due to the LT precession, $\mathbf{\Omega_{LT}\times L}$ (Scheuer \& Feiler 1996; Lodato \& Pringle 2006):
\begin{eqnarray}
\frac{\partial \mathbf{L}}{\partial t} =&& \frac{3}{R} \frac{\partial}{\partial R} [\frac{R^{1/2}}{\Sigma} \frac{\partial}{\partial R} (\nu_1 \Sigma R^{1/2}) \mathbf{L}] \nonumber \\
&& +\frac{1}{R} \frac{\partial}{\partial R} [(\nu_2 R^2 |\frac{\partial \mathbf{l}}{\partial R}|^2 -\frac{3}{2} \nu_1) \mathbf{L}] \nonumber \\
&&  +\frac{1}{R} \frac{\partial}{\partial R}(\frac{1}{2} \nu_2 R L \frac{\partial \mathbf{l}}{\partial R}) + \mathbf{\Omega_{LT}\times L},
\end{eqnarray}
where $L=|\mathbf{L}|=\Sigma \sqrt{GM_{\bullet}R}$ is the angular momentum per unit area of disk, $\Sigma$ is the surface density, $\mathbf{l}$ is the unit vector indicating the local direction of the specific angular momentum in the disk. Note that $R$ is a spherical coordinate, and is not the cylindrical radius. 

For the thin disk case (i.e. $H/R<\alpha$), warp propagation would occur in a diffusive way (Popatoizou \& Lin 1995). The transition radius between the two regimes is known as the Bardeen-Petterson radius $R_{\rm{BP}}$, which is the radius at which the warping propagation time-scale $t_{\nu_2}=R^2/\nu_2$ equals the local forced precession rate $\tau_{\rm p}$, i.e. (Scheuer \& Feiler 1996; Caproni et al. 2007) 
\begin{equation}
R_{\rm{BP}} = \sqrt{\frac{\nu_2}{\Omega_{\rm{LT}}(R_{\rm{BP}})}}.
\label{RBP}
\end{equation}
Here again $\nu_2$ is the viscosity normal to the accretion disk. Roughly speaking, the disk becomes aligned with the BH spin at $R\ll R_{\rm{BP}}$ and keeps its original inclination for $R\gg R_{\rm{BP}}$. 

Scheuer \& Feiler (1996) have found an analytical steady-state solution to the above equation (removing the time-dependent term $\partial \mathbf{L}/\partial t$), subject to various simplifying assumptions about the disk density distribution and the inner boundary condition. This solution may be inaccurate at very small radii from the BH. It is based on a first-order approximation and a small inclination angle ($\theta_{\rm{orbit}} \ll 1$) assumption. In our modelling, instead of applying this solution, we solve the equations numerically following the numerical scheme outlined in Pringle (1992).

A small-amplitude warp propagates diffusively in the linear regime. In this regime, there is a relation between $\nu_1$ and $\nu_2$, which is obtained analytically by Ogilvie (1999) and further confirmed numerically by Lodato \& Price (2010). This relation reads 
\begin{equation}
\frac{\nu_2}{\nu_1} \equiv f(\alpha)= \frac{1}{2\alpha^2}
\frac{4(1+7\alpha^2)}{4+\alpha^2}. 
\label{nu2/nu1}
\end{equation}
In the non-linear case (warps with large amplitudes), additional dissipation caused by fluid instability might reduce $f(\alpha)$. One would have $\nu_2/\nu_1 \simeq 1$ (Gammie, Goodman \& Ogilvie 2000). In our calculations, we adopt $\alpha=0.2$ and assumes that the disk is in the linear regime, so that Eq.(\ref{nu2/nu1}) is valid. 

The solution for the inclination angle at time $t$ and radius $R$ depends on the following parameters: the BP radius $R_{\rm{BP}}$, the outer disk boundary radius $R_{\rm{out}}$, and the orbital inclination $\theta_{\rm{orbit}}$. In our calculation, we feed matter at a constant rate with a constant inclination angle $\theta_{\rm{orbit}}$ to the grid near the outer boundary $R_{\rm{out}}$. We use 50 grid points spaced linearly between $R_{\rm ms}$ and $R_{\rm out}$. The time unit is $t_0= R_{\rm g}^2/\nu_2 = R_{\rm BP}/2 a_\bullet c \sim 5~{\rm s}~(R_{\rm BP}/R_{\rm g})/(2a_\bullet) (M_\bullet/10^6M_\odot)$, which is $\sim 28 \rm s$ for $R_{\rm BP}/R_{\rm g} = 10$, $a_\bullet=0.9$ and $M_\bullet = 10^6 M_\odot$. Following Pringle (1992), we take a time-step  $dt = 10^{-3} t_0$. We run the code for about $10^5$ time-steps without the LT effect to settle the disk into a numerically equilibrium state. We then evolve the disk by applying the forced precession due to the LT torque. The disk starts to warp quickly, but the twisting evolution soon slows down. It takes about $2\times 10^6$ runs when the inclination no longer evolves significantly, and we take this as the steady state. This corresponds to $\sim 0.6$ days, which is shorter than the fall back time scale ($\sim 5$ days) and the Lense-Thirring period ($\sim 2.7$ day). We therefore treat the disk as in a steady state\footnote{In principle, one should also consider the long-term evolution of the disk that aligns the disk normal with the black hole spin axis. The time scale for such alignment can be estimated as (Lodato \& Pringle 2006) $t_{\rm align}=3 a_\bullet (\nu_1/\nu_2) (M_\bullet/\dot{M}) (R_{\rm g}/R_{\rm BP})^{1/2} \sim 10^6$ yr, which is much longer than the time scales studied in this paper. We therefore ignore the alignment effect, and treat the disk as a steady state.} 

\begin{figure}[htbp]
\centerline{\includegraphics[width=70mm,height=70mm]{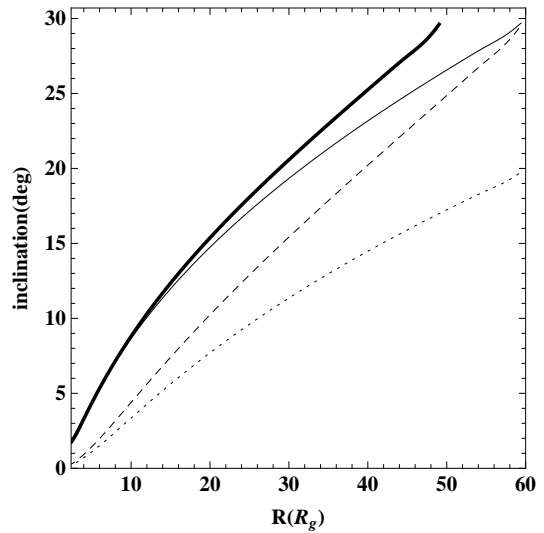}}
\caption[]{Disk inclination $\theta_{\rm{inc}}$ vs. radius $R$ for different $R_{\rm{BP}}$, $R_{\rm{out}}$ and $\theta_{\rm{orbit}}$: Thick solid line: $R_{\rm{BP}}=10 R_{\rm{g}}$, $R_{\rm{out}}=50 R_{\rm{g}}$ and $\theta_{\rm{orbit}}=30^{\circ}$; thin solid line: $R_{\rm{BP}}=10 R_{\rm{g}}$, $R_{\rm{out}}=60 R_{\rm{g}}$ and $\theta_{\rm{orbit}}=30^{\circ}$; dashed line: $R_{\rm{BP}}=25 R_{\rm{g}}$, $R_{\rm{out}}=60 R_{\rm{g}}$ and $\theta_{\rm{orbit}}=30^{\circ}$; and dotted line: $R_{\rm{BP}}=25 R_{\rm{g}}$, $R_{\rm{out}}=60 R_{\rm{g}}$ and $\theta_{\rm{orbit}}=20^{\circ}$. } \label{fig3}
\end{figure}

In Fig.\ref{fig3}, the numerical solutions in the steady state are shown for different $R_{\rm{BP}}$, $R_{\rm{out}}$ and $\theta_{\rm{orbit}}$. We find that the inclination of the inner disk critically depends on the BP radius $R_{\rm{BP}}$ and the initial inclination of the stellar orbit $\theta_{\rm{orbit}}$. The effect of the outer disk boundary is not significant for the inner disk.

A precessing disk would induce jet precession. In our model, the relativistic jet is launched near the BH by the Blandford-Znajek process (Lei \& Zhang 2011). However, the jet direction is not determined by the normal direction at the inner disk, but is rather defined by normal direction at an outer region in the disk near $R_{_{\rm BP}}$. This is because the magnetic field threading the disk would drive an outflow from the disk surface via the Blandford \& Payne (1982) mechanism. A centrifugally driven outflow is possible if the poloidal component of the magnetic field makes an angle less than $60 ^{\circ}$ with the disk surface. This wind would collimate the BZ jet, making it precess with the angle and period defined from the launching site. This launching site is most likely at $R_{_{\rm BP}}$, where the disk inclination angle significantly changes. A similar proposal was introduced to study other BH accretion systems (e.g. Begelman et al. 2006) and gamma-ray bursts (e.g. Lei et al. 2007). In the following calculation, we suggest that the precession period and inclination angle of the relativistic jet is defined by the physical conditions at $R_{\rm{BP}}$.

\section{Observational Constraints}

Some constraints on the model can be derived from the observational data outlined in Section 2. 

\begin{figure}[htc]
\center
\includegraphics[width=7cm,angle=0]{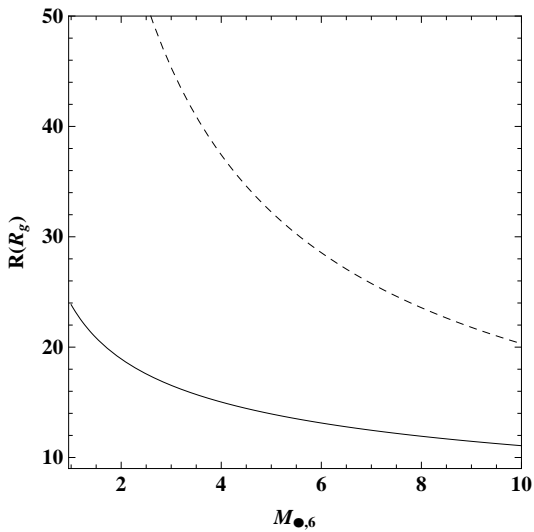}
\caption{The constraint on the Bardeen-Petterson radius $R_{\rm{BP}}$ (solid line) as a function of BH mass, based on the requirement of $\tau_{\rm{p}}(R_{\rm{BP}})=2.7$days with $a_{\bullet}=0.9$. The disk outer edge $R_{\rm{out}}$ is also shown as the dashed line.}
\label{fig4}
\end{figure}

The first constraint is the 2.7-day quasi-period. In the precession model delineated above, this period should correspond to the LT period at the BP radius, $R_{\rm{BP}}$. According to Lei \& Zhang (2011), the most probable value of the BH spin in Sw J1644+57 is $a_{\bullet}=0.9$. We take this value for the BH spin in the calculation. Applying $\tau_{\rm{p}}=2.7$days into Eq.(\ref{eq_taup}), we can estimate the BP radius $R_{\rm{BP}}$ that satisfies the observations (Fig.\ref{fig4}). For comparison, we also show the disk outer edge $R_{\rm{out}}$ in Fig.\ref{fig4}, where $M_*=M_{\sun}$, $R_*=R_{\sun}$, $\kappa=1$ and $\eta=2$ are adopted as typical values. It is found that $R_{\rm{BP}}$ is located close to the BH. We note that having a small $R_{\rm BP}$ to allow a 2.7-day LT precession period is consistent with the theoretical expectation of the accretion model. Applying equations (\ref{RBP}) and (\ref{nu2/nu1}), one can get $\nu_1 \sim 4\alpha^2 (a_\bullet R_{\rm g}/R_{\rm BP}) (GM_{\bullet}/c)$. Using the $\alpha$ prescription, i.e. $\nu_1 = \alpha C_{\rm s} H$ (here $C_{\rm s}$ is the sound speed), and applying the thin-disk condition $H/R < \alpha$, one can get the requirement of $\alpha > 0.1$ for $R_{\rm BP}/R_{\rm g} = 10$ and $a_\bullet = 0.9$, which is satisfied for the conventional parameterization of $\alpha$.

The next observational constraint comes from the observed shape and depth of the dips seen in the lightcurve. They are defined by the complicated combinations of several unknown parameters, including the bulk Lorentz factor $\Gamma$, the jet half opening angle $\theta_{\rm j}$, the inclination angle of the jet $\theta_{\rm i}$, as well as the observer's viewing angle $\theta_{\rm obs}$ with respect to the BH spin axis. In the following we discuss these constraints in detail.

At any time $t$, the angle between the observer $\hat {r}_{\rm{obs}}$ and jet axis $\hat {r}_{\rm{jet}}(t)$ (the axis of the presumed conical jet) is denoted by $\psi(t)$ (see Fig.\ref{fig2}), which is defined as
\begin{equation}
\label{eq18}
\psi(t) = \cos ^{ -1}(\hat {r}_{\rm{obs}} \cdot \hat {r}_{\rm{jet}}(t) ).
\label{psit}
\end{equation}
A dip in the lightcurve can be seen if the line of sight is outside the jet cone, i.e., $\psi(t)>\theta_{\rm{j}}$.

For simplicity, we start with a uniform conical jet, so that when $\psi<\theta_{\rm{j}}$ the observer receives a uniform flux. For an off-axis observer at $\psi>\theta_{\rm{j}}$, the observed flux density can be written as (e.g. Granot et al. 2002)
\begin{equation}
F_{\nu}(\psi,t) = D^3 F_{\nu /D}(0,D t),
\end{equation}
where $D=(1-\beta)/(1-\beta \cos \psi)$, $\beta= \sqrt{1-1/\Gamma^2}$ and $\psi=\rm{max}(0,\theta_{obs}-\theta_j)$. Define $F(\psi,t)=\nu F_{\nu}(\psi,t)$, we have the X-ray flux at $\psi$
\begin{equation}
F(\psi,t) = D^4 F(0,D t).
\end{equation}
For a tidal disruption event, the long-term lightcurve is defined by the fall-back accretion rate, which is $\propto t^{-5/3}$. The long term X-ray lightcurve of Sw J1644+57 seems to be consistent with this behavior, i.e. $F(t) \propto t^{-5/3}$. We therefore have
\begin{equation}
F(\psi,t) = D^{7/3} F(0,t).
\end{equation}

\begin{figure}[htbp]
\centerline{\includegraphics[width=70mm,height=70mm]{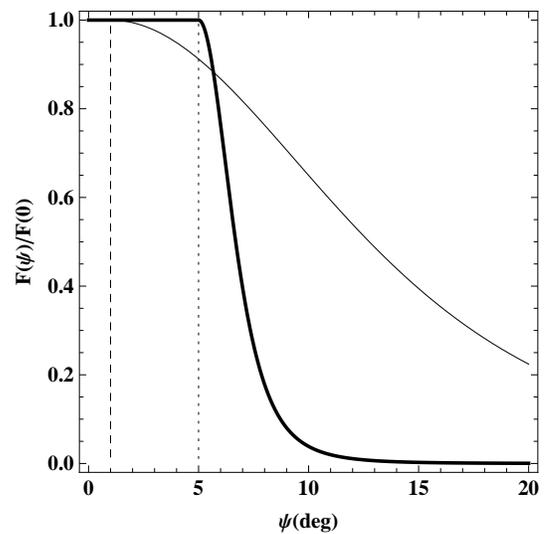}}
\caption[]{The flux as a function of view angle $\psi$ for the jet for two case: $\Gamma=3$ and $\theta_{\rm{j}} = 1^{\circ}$ (thin line);  and $\Gamma=20$ and $\theta_{\rm{j}} = 5^{\circ}$ (thick line).} \label{fig5}
\end{figure}

The shape the lightcurve depends on several factors. The first one is the Lorentz factor. In Fig.\ref{fig5} we show how the Lorentz factor affects the observed structure of jet. Even if the physical jet has a sharp cutoff near the edge, the observed jet structure has a wing defined by the gradual decrease of the Doppler factor. If the jet Lorentz factor is large, the flux drops sharply as the line of sight goes away from the jet cone, and one would see a very deep dip. On the other hand, the dip would be shallower if the jet has a smaller Lorentz factor. In the observed lightcurve, the dips do not have a uniform depth, and in some periods the dips disappear completely. This might be caused by fluctuation of the jet Lorentz factor as a function of time. 

The second parameter to shape the peaks and dips in the lightcurve is the duty cycle 
\begin{equation}
\chi \equiv t_{\rm{obs}}/\tau_{\rm{p}} < 1,
\label{eq_chi}
\end{equation}
where $t_{\rm{obs}}$ is the time when the observer's line of sight is inside the jet cone during one precession period. A large value of $\chi$ would correspond to narrow dips, while a small $\chi$ would correspond to wide dips. This ratio can be obtained by solving the equation $\psi(t)=\theta_{\rm{j}}$. In general one obtains $\chi=\chi(\theta_{\rm{i}},\theta_{\rm{obs}},\theta_{\rm{j}})$.

The last parameter 
\begin{equation}
\lambda (\theta_{\rm{i}},\theta_{\rm{obs}},\theta_{\rm{j}}, \Gamma) \equiv \frac{F(\psi_{\rm{max}})}{F(0)} = \frac{F(\theta_{\rm{i}}+\theta_{\rm{obs}})}{F(0)} < 1,
\label{eq_lambda}
\end{equation}
defines how deep the dip is. Besides the three angle parameters $(\theta_{\rm{i}},\theta_{\rm{obs}},\theta_{\rm{j}})$, the Lorentz factor $\Gamma$ also enters the problem. 

Finally, the jet opening angle $\theta_{\rm{j}}$ and the mean Lorentz factor $\Gamma$ of Sw J1644+57-like events can be constrained through a statistical argument. Based on the comparison between observed event rate of these events and the more general TDE event rate constrained from both TDE theory and observational data, one can constrain the beaming factor $f_{\rm{b}}$ (Burrows et al. 2011; Lei \& Zhang 2011) 
\begin{equation}
f_{\rm{b}} \sim \rm{max}(\frac{1}{2\Gamma^2}, \frac{\theta_j^2}{2}) \sim  \frac{{\cal R}_{\rm obs}}{10\% {\cal R}_{\rm tot}} \leq 5.5 \times 10^{-3}.
\end{equation}
This requires that $\Gamma \geq 9.5$ and $\theta_{\rm{j}} \leq 6^{\circ}$. In the following calculations, we take $\theta_{\rm{j}}=6^{\circ}$ and $\Gamma=20$ as the typical values. A relative large value of $\Gamma$ is demanded to reproduce the depth of the observed dips.

\begin{figure}[htbp]
\centerline{\includegraphics[width=70mm,height=70mm]{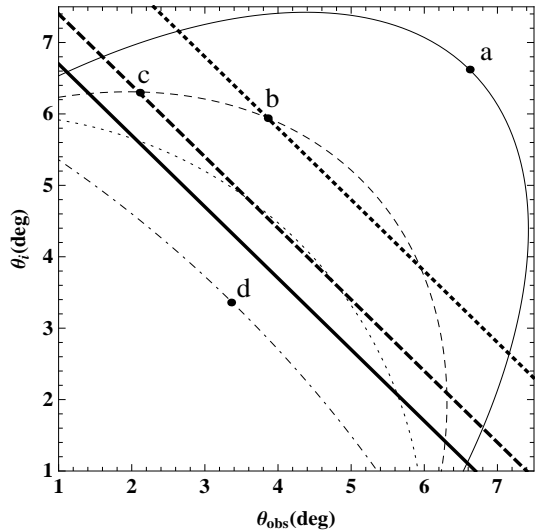}}
\caption[]{The relation between $\theta_{\rm{i}}$ and $\theta_{\rm{obs}}$ for different $\chi$ (thin lines, solid: $\chi=0.3$; dashed: $\chi=0.4$; dotted: $\chi=0.5$; dot-dashed: $\chi=0.7$) and $\lambda$ (thick lines, solid: $\lambda=0.5$; dashed: $\lambda=0.3$; dotted: $\lambda=0.1$), where the jet opening angle $\theta_{\rm{j}}= 6^{\circ}$ and the Lorentz factor $\Gamma = 20$.} 
\label{fig6}
\end{figure}

\begin{figure}[htbp]
\center
\includegraphics[width=6cm,angle=0]{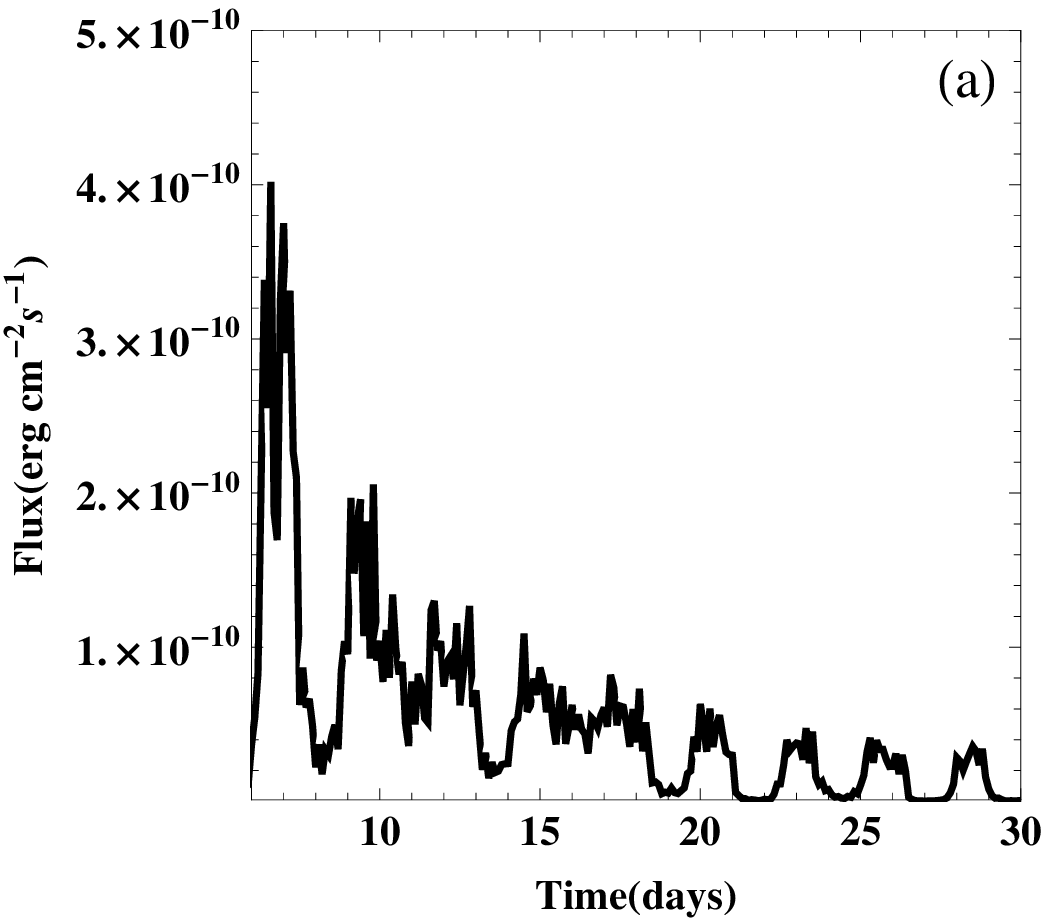}
\includegraphics[width=6cm,angle=0]{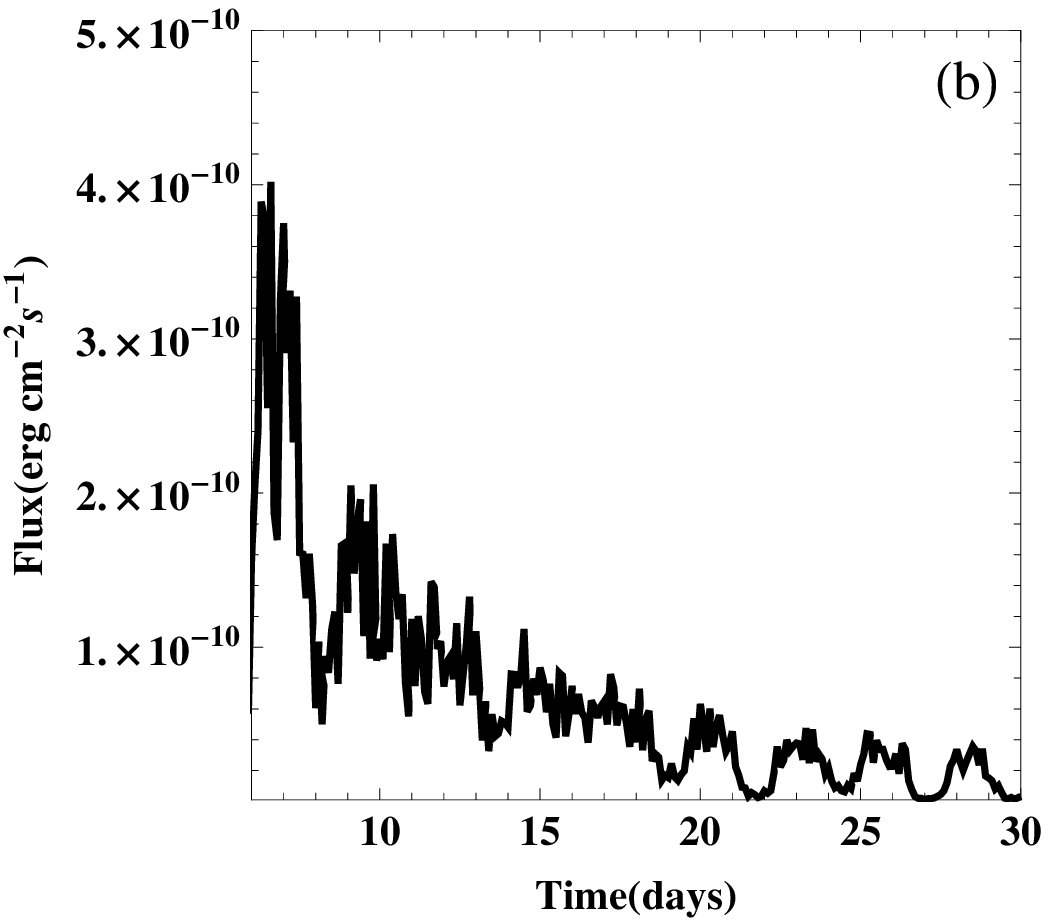}
\includegraphics[width=6cm,angle=0]{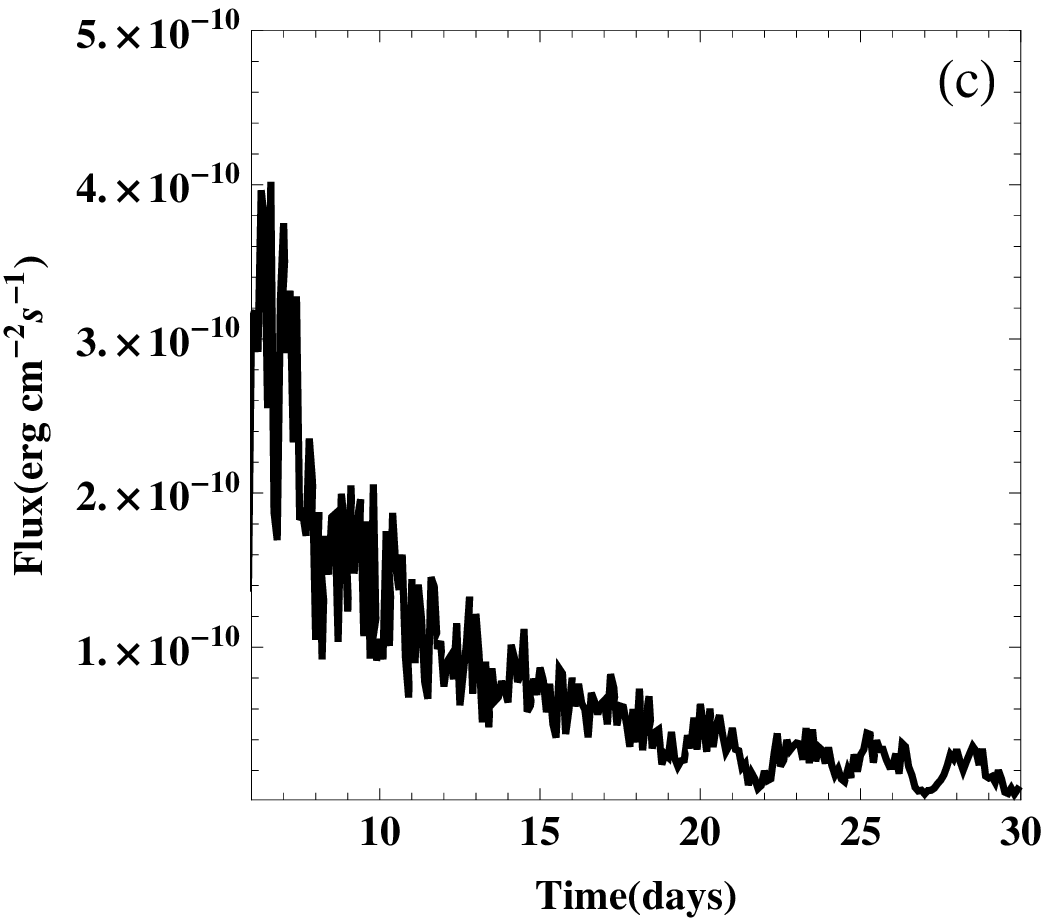}
\includegraphics[width=6cm,angle=0]{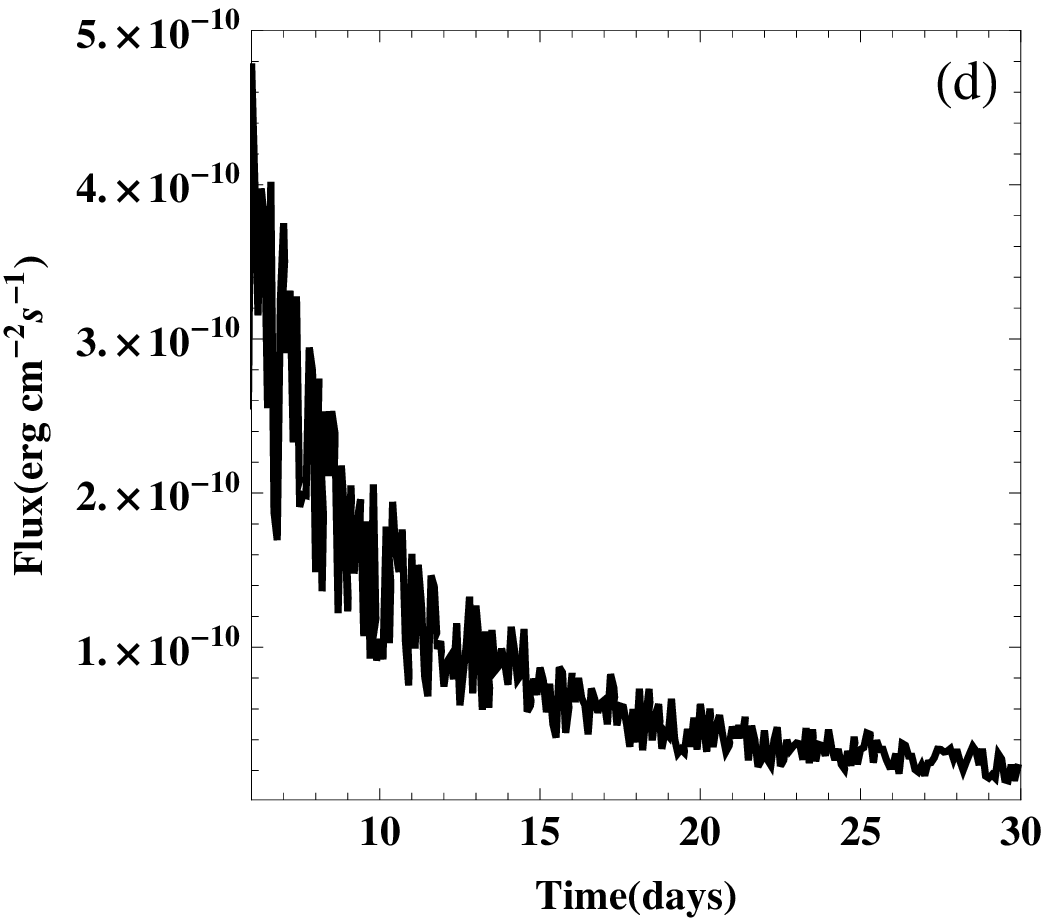}
\caption{Several simulated XRT lightcurves of Sw J1644+57. The four cases correspond to the ``a,b,c,d'' points of the parameter space in Fig. \ref{fig6}. (a): $\theta_{\rm{obs}}=5.0$ deg, $\theta_{\rm{i}}=7.4$ deg; (b): $\theta_{\rm{obs}}=4.0$ deg, $\theta_{\rm{i}}=6.0$ deg; (c): $\theta_{\rm{obs}}=2.1$ deg, $\theta_{\rm{i}}=6.3$ deg; (d): $\theta_{\rm{obs}}=3.3$ deg, $\theta_{\rm{i}}=3.3$ deg. It is shown that the case (a) has too sharp dips, while the case (d) essentially smeared out the dips. The cases (b) and (c) would be close to what is observed from the source.}
\label{fig7}
\end{figure}

The values of $\theta_{\rm{i}}$ and $\theta_{\rm{obs}}$ can be then obtained by combining Eqs. (\ref{eq_chi}) and (\ref{eq_lambda}), if we obtain information about the values of $\chi$ and $\lambda$. In Fig.\ref{fig6}, the relation between $\theta_{\rm{i}}$ and $\theta_{\rm{obs}}$ for different $\chi$ and $\lambda$ are shown. Based on observations, the value of $\chi$ in Sw J1644+57 cannot be too large (e.g., point ``d'' in Fig.\ref{fig6}), since otherwise the dips would be too shallow to be observed (e.g. Fig.\ref{fig7}d). Similarly, $\lambda$ cannot be too small (e.g., point ``a'' in Fig.\ref{fig6}), since otherwise the peaks and dips would be too sharp (e.g. Fig.\ref{fig7}a). Observationally, moderate values of both parameters, e.g. $\chi \sim 0.4$ and $\lambda \sim 0.3$, are favored (e.g. points ``b'' or ``c'' in Fig.\ref{fig6}, see Figs.\ref{fig7}b and \ref{fig7}c). Since we observe most of the jet beam, the line of sight cannot be too far away from the BH spin axis, i.e. $\theta_{\rm{obs}}$ is expected to be small. From Fig.\ref{fig6}, these constraints suggest that the jet inclination angle $\theta_{\rm{i}}$ is likely similar to $\sim \theta_{\rm{j}}$.

\begin{figure}[htbp]
\centerline{\includegraphics[width=70mm,height=70mm]{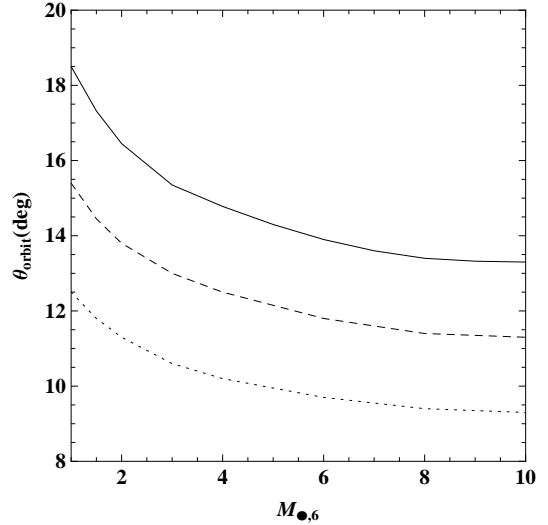}}
\caption[]{The inferred inclination of stellar orbit $\theta_{\rm{orbit}}$ vs. the BH mass, where $\tau_{\rm{p}}=2.7$ days. The solid, dashed and dotted lines correspond to $\theta_{\rm{i}}=7^{\circ}$, $6^{\circ}$ and $5^{\circ}$, respectively. } 
\label{fig8}
\end{figure}

Based on the above discussion, the BP radius $R_{\rm{BP}}$ could be obtained with the observed jet precession period (2.7 days). Putting all the constrains into the numerical model described in Section 2, we can infer the inclination of the stellar orbit before disruption $\theta_{\rm{orbit}}$. Fig.\ref{fig8} shows the inferred inclination of the stellar orbit $\theta_{\rm{orbit}}$ as a function of black hole mass. Different lines correspond to different $\theta_{\rm i}$ values. The calculation was proceeded by re-running the numerical disk model for each $M_{\bullet,6}$ and $\theta_{\rm i}$, until the proper $\theta_{\rm{orbit}}$ is found for the required inclination angle at $R_{\rm BP}$. From Fig.\ref{fig8} we find that the inclination angle of the initial stellar orbit can be as large as $18^{\circ}$. For small BH masses, $\theta_{\rm{orbit}}$ can be much greater than $\theta_{\rm{i}}$. However, for large BH masses, $\theta_{\rm{orbit}}$ is not much larger than $\theta_{\rm{i}}$, mainly due to a small value of $R_{\rm{T}}/R_{\rm g}$ involved. 

Finally, the observed X-ray lightcurve Sw J1644+57 does not have strict periodic dips. In some phases the dips disappear, and in some phases the dips do not appear at exactly the predicted phase (Burrows et al. 2011, and Fig.\ref{fig1}). We believe that this is related to some stochastic processes involved in accretion and jet launching. The Lorentz factor $\Gamma$ may vary with time, and $R_{_{\rm BP}}$ may slightly vary at different epochs. In order to test this idea,  we carry out a range of simulations to the lightcurves. First, we introduce a small time-scale variability in the lightcurve overlapped on the $F(0,t)\propto t^{-5/3}$ envelope (Burrows et al. 2011), which may be a result of stochastic magnetic dissipation (e.g. Zhang \& Yan 2011) or fluctuations of accretion rate (e.g. Wang et al. 2006). For simplicity, the function $\psi(t)$ (\ref{psit}) is taken as an oscillatory function of $t$ with frequency equal to the LT frequency. Next, we introduce a variation of Lorentz factor by allowing $\Gamma$ to randomly vary in the range of (2, 25) during each precession period. We simulate four cases in Fig.\ref{fig7}, which correspond to ``a'', ``b'', ``c'' and ``d'' in Fig.\ref{fig6}. We find that the change of Lorentz factor can indeed result in the disappearance of dips at some epochs. The peaks and dips in Fig.\ref{fig7}a are well separated and are too spiky. The lightcurve in Fig.\ref{fig7}d on the other hand is too smooth, without clear evidence of dips. This suggests that they do not resemble the geometric configuration of Sw J1644+57. The other two cases, i.e., Figs.\ref{fig7}b and \ref{fig7}c, give interesting quasi-periodic features with noticeable dips. Even though we cannot fully constrain the geometry of the Sw J1644+57 system, we can say that it should resemble the conditions of these two cases based on the simulation results. In our calculation, a relatively large $\Gamma$ is needed to give significant dips. Late time X-ray afterglow months after trigger no longer show significant dips with a 2.7-day period (D. N. Burrows, 2012, private communication). This could be understood as due to a systematically reduction of $\Gamma$ at late epochs when luminosity drops significantly.

\section{Conclusions and Discussion}

In this paper, we propose that the Sw J1644+57 jet launched from a SMBH precesses. This comes from two independent arguments. First, a tidally disrupted star should not have knowledge about the spin direction of the BH before disruption, so that the original stellar orbit very likely has an inclination with respect to the equatorial plane of the BH. A rapidly spinning BH (which is probably the main reason to launch a relativistic jet, Lei \& Zhang 2011) tends to distort the accretion flow and warp the accretion disk due to strong LT frame dragging. This would naturally result in a precessing jet if the accretion rate is low enough. Second, observationally two pieces of evidence hint towards a precessing jet: the rough 2.7-day periodicity and the noticeable lightcurve dips. We performed a SFC analysis to the observed 5-30 day X-ray lightcurve, and confirmed the 2.7-day rough period. We then carried out detailed theoretical modeling on disk warping and jet precessing, and inferred the underlying parameters of the jet system from the observational data. We found that due to disk warping, the original stellar orbit does not have to be nearly aligned with the BH equatorial plane (cf. Stone \& Loeb 2012). Rather, $\theta_{\rm orbit}$ can be as large as $18^{\circ}$. In order to reproduce the depths of the lightcurve dips and their distribution, the Lorentz factor should range from moderate values to values as high as 20. The duty cycle $\chi$, which describes the fraction of time when the line of sight intersects the jet, is found to be a moderate value, slightly smaller than 0.5.

Through simulations, we were able to reproduce lightcurves similar to the observed one. Saxton et al. (2012) interprets the diminishing dips in some periods as due to jet nutation. Our SFC analysis does not show another quasi frequency that might be related to nutation. We attribute the diminishing dips to fluctuations in $\Gamma$. The mis-match between the 2.7-day period and some dips is also understandable, since the outflow that collimates the jet may not always launched strictly from $R_{\rm{BP}}$, or $R_{\rm{BP}}$ could vary stochastically during the event.

A warp is also possible to propagate in a wave-like mode when $\alpha \ll 1$ or when the disk is rather thick (Papaloizou \& Lin 1995, Papaloizou \& Terquem 1995, Lubow \& Ogilvie 2000). Lubow, Ogilvie \& Pringle (2002) found the solutions of wave-like warped disk. For a steady-state prograde disk, the inner disk still has a significant inclination angle. However, our modeling suggests that in order to reproduce data, a relatively small inclination angle at the inner disk is required. This model is therefore not favored. A steady-state retrograde on the other hand has a small-inclination angle at inner radii, which could be consistent with the data. Since it is entirely possible that the incoming star comes from a direction to make a retrograde disk, and since the BZ jet power does not sensitively depend on the orientation of the disk spin with respect to the black hole spin (Tchekhovskoy et al. 2012), we believe that a retrograde disk with wave-like warping (Lubow et al. 2002) would be another possibility to interpret the phenomenology of Sw J1644+57.

One major uncertainty in our model is viscosity. In our modelling, we have adopted $\alpha=0.2$ throughout, and used Eq.(\ref{nu2/nu1}) to calculate the ratio $\nu_2/\nu_1$. Varying $\alpha$ within a reasonable range ($\alpha > 0.1$ required by the thin disk condition) would not significantly change our results, since the numerical solutions are only sensitive to $R_{\rm BP}$ when $R_{\rm out}$ and $\theta_{\rm orbit}$ are given, while $R_{\rm BP}$ is constrained by the observed 2.7-day quasi-period.

\acknowledgements We thank the referee for constructive comments,  D. N. Burrows, R.-F. Shen, R. Soria, C. Saxton, K. Wu,  P. Kuin, and other members of UNLV GRB group for helpful discussion, and A. Maxham for technical assist. This work is supported by NSF under Grant No. AST-0908362, by National Natural Science Foundation of China (grants 11003004, 11173011 and U1231101), and National Basic Research Program (``973'' Program) of China under Grant No. 2009CB824800. BZ acknowledges a Cheung Kong scholarship for support. WHL acknowledges a Fellowship from China Scholarship Program for support.











\end{document}